\documentclass{jetpl}
\usepackage {cite}
\twocolumn

\lat

\title{Stability of the superconducting $d_{x^2-y^2}$-wave pairing
towards the intersite Coulomb repulsion between oxygen holes in
high-T$_c$ superconductors}

\rtitle{Stability of the superconducting $d_{x^2-y^2}$-wave
pairing towards the intersite Coulomb repulsion...}

\sodtitle{Stability of the superconducting $d_{x^2-y^2}$-wave
pairing towards the intersite Coulomb repulsion between oxygen
holes in high-T$_c$ superconductors}

\author{V.\,V.\,Val'kov$^{a}$,
D.\,M.\,Dzebisashvili$^{a,b}$, M.\,M.\,Korovushkin$^{a}$, and
A.\,F. Barabanov$^{c}$}

\rauthor{V.\,V.\,Val'kov, D.\,M.\,Dzebisashvili,
M.\,M.\,Korovushkin, and A.\,F. Barabanov}

\address{$^{a}$Kirensky Institute of Physics, 660036 Krasnoyarsk, Russia \\
$^{b}$M.\,F. Reshetnev Siberian State Aerospace University, 660014
Krasnoyarsk, Russia\\
$^{c}$Institute for High Pressure Physics, 142190 Troitsk, Moscow
Region, Russia}

\abstract{It is shown that an account for the space separatedness
of the two-orbital subsystem of the oxygen holes and the subsystem
of the localized spins of copper ions in high-T$_c$ cuprate
superconductors leads to the stability of the superconducting
$d_{x^2-y^2}$-wave pairing towards the strong Coulomb repulsion
between holes located at the nearest oxygen ions. This effect is
due to the fact that the Coulomb potential slips out of the
equation for the Cooper pairing in the $d_{x^2-y^2}$-wave channel
owing to the properties of symmetry.}


\begin{document}

\maketitle

{\bf 1. Introduction}

It is known that the Cooper pairing of the fermions resulting from
the kinematic~\cite{Zaitsev87}, exchange and
spin-fluctuation~\cite{Izyumov9799,Plakida10} mechanisms, which
are considered in the framework of the Hubbard
model~\cite{Zaitsev8889,Zaitsev04,Val'kov11}, $t-J$
model~\cite{Izyumov9799,Plakida10,Kagan94}, and $t-J^*$
model~\cite{Yushankhai90,Val'kov02,Val'kov03}, is suppressed with
regard to the intersite Coulomb interaction $V$ between the
carriers located at the nearest sites of the lattice. This effect
manifests itself most strongly in the $d$-wave
channel~\cite{Plakida13}, so at the values
$V\sim1-2~\mathrm{eV}$~\cite{Fischer11} the Cooper instability is
turned out to be totally suppressed. The $s$-wave pairing caused
by the stronger kinematic mechanism~\cite{Zaitsev87} is more
stable and, as a result, the Cooper pairing is robust even with
regard to $V$~\cite{Zaitsev8889,Zaitsev04,Val'kov11}. Thus, the
contradiction between the theoretical and experimental results
arises: an account for the Coulomb repulsion leads to the
suppression of the $d$-wave pairing, which is experimentally
observed, but maintains the $s$-wave pairing, which is not
realized in the experiment. This fact significantly restricts the
capabilities of the above-mentioned theories of high-T$_c$
superconductors.

In this paper, it is shown that an account for the real structure
of CuO$_2$ plane in the framework of the Emery
model~\cite{Emery87,Varma87} removes the above-mentioned
contradiction. In our theory, the Fourier transform corresponding
to the Coulomb repulsion between the holes located at the nearest
oxygen sites slips out of the equation for the Cooper pairing in
the $d$-wave channel owing to the properties of symmetry. At the
same time, the self-consistent equation for the $s$-wave pairing
contains the contribution of the Coulomb interaction and, as a
result, this pairing is suppressed. Thus, our theory answers the
question of why the superconducting $d$-wave pairing survives with
regard to the intersite Coulomb repulsion between oxygen holes, as
well as why the $d_{x^2-y^2}$-wave pairing instead of the $s$-wave
pairing occurs in cuprate superconductors. \vspace{8pt}

{\bf 2. The spin-fermion model}

In the regime of the strong electron correlations, in which the
Hubbard repulsion between holes $U_d$ is large, i.e.,
$U_d>\Delta_{pd}\gg t_{pd}$, the Emery model can be reduced to the
spin-fermion model~\cite{Zaanen88,Barabanov97a} with the
Hamiltonian
\begin{eqnarray}\label{Hamiltonianspinfermion}
&&\hat{H}=\hat{H}_0+\hat{J}+\hat{V}+\hat{I},
\end{eqnarray}
$$ \hat{H}_0=\sum_{k\alpha}\Bigl(\xi_0(k_x)
a_{k\alpha}^{\dagger}a_{k\alpha}+ \xi_0(k_y)
b_{k\alpha}^{\dagger}b_{k\alpha}+$$
$$ \qquad+t_{k}(
a_{k\alpha}^{\dagger}b_{k\alpha}+
b_{k\alpha}^{\dagger}a_{k\alpha})\Bigr),
$$
$$
\hat{J}=\frac{J}{N}\sum_{\substack{fkq\\\alpha\beta}}e^{if(q-k)}
u_{k\alpha}^{\dag}(\textbf{S}_f\boldsymbol{\sigma}_{\alpha\beta})u_{q\beta},
$$
$$
\hat{V}=V\sum_{f\Delta}\hat{n}_{f+\frac{x}{2}}\hat{n}_{f+\frac{x}{2}+\Delta},
\qquad\hat{I}=\frac{I}{2}\sum_{\langle
fm\rangle}\textbf{S}_f\textbf{S}_m,
$$
which describes the subsystem of the oxygen holes interacting with
the localized spins in copper ions. Here
\begin{eqnarray}\label{Hamiltonian_notations}
&&\xi_0(k_{x(y)})=\varepsilon_p-\mu+\tau(1+\cos
k_{x(y)}),\nonumber\\
&&t_{k}=(2\tau-4t)\cos\frac{k_x}{2}\cos\frac{k_y}{2},\nonumber\\
&&u_{k\beta}=\cos\frac{k_x}{2}a_{k\beta}+
\cos\frac{k_y}{2}b_{k\beta},\nonumber\\
&&\tau=\frac{t_{pd}^2}{\Delta_{pd}}
\biggl(1-\frac{\Delta_{pd}}{U_d-\Delta_{pd}-2V_{pd}}\biggr),\nonumber\\
&&J=\frac{4t_{pd}^2}{\Delta_{pd}}
\biggl(1+\frac{\Delta_{pd}}{U_d-\Delta_{pd}-2V_{pd}}\biggr),\nonumber\\
&&I=\frac{4t_{pd}^4}{(\Delta_{pd}+V_{pd})^2}\biggl(\frac{1}{U_d}+\frac{2}{2\Delta_{pd}+U_p}\biggr).
\end{eqnarray}
The Hamiltonian $\hat{H}_0$ describes the subsystem of oxygen
holes in the momentum representation. The operators
$a_{k\alpha}^{\dagger}(a_{k\alpha})$ create (annihilate) the holes
with spin $\alpha=\pm1/2$ in the oxygen subsystem with the
$p_x$-orbitals. Similarly, the operators
$b_{k\alpha}^{\dagger}(b_{k\alpha})$ operate in the oxygen
subsystem with the $p_y$-orbitals. The initial one-site energy of
holes is $\varepsilon_p$; $\mu$ is the chemical potential; $t$ is
hopping parameter of the oxygen holes. The exchange interaction
between the oxygen subsystem and the subsystem of the localized
spins is described by the operator $\hat{J}$. Here, $\textbf{S}_f$
is the vector operator of the spin moment on the copper ion in the
site with index $f$ and
$\boldsymbol{\sigma}=(\sigma^x,\,\sigma^y,\,\sigma^z)$ is the
vector that consists of the Pauli matrices. The Coulomb
interaction between holes located at the nearest oxygen ions is
described by the operator $\hat{V}$. Here,
$\hat{n}_{f+x(y)/2}=\sum_{\sigma}\hat{n}_{f+x(y)/2,\sigma}$ is the
operator of the number of holes at the oxygen ion with index
$f+x(y)/2$ ; $\textbf{x}=(1,0)$ and $\textbf{y}=(0,1)$ are lattice
basis vectors in the units of the lattice parameter; $\Delta$ is a
vector connecting the nearest oxygen ions. The last term of the
Hamiltonian describes the exchange interaction between the nearest
spins of copper ions. The intensity of this interaction is giving
by the parameter $I$.

Below we use the following commonly accepted parameter values:
$t_{pd}=1.3\,\textrm{eV}$, $\Delta_{pd}=3.6\,\textrm{eV}$,
$U_d=10.5\,\textrm{eV}$, $V_{pd}=1.2\,\textrm{eV}$,
$V=1-2\,\textrm{eV}$~\cite{Hybertsen89,Ogata08,Fischer11}. For
these values the parameter of exchange interaction
$I~=~0.136\,\textrm{eV}\,(1570\,\textrm{K})$ agrees well with the
available experimental data~\cite{Ogata08}. For the hopping
parameter of the oxygen holes, we use the value
$t=0.1\,\textrm{eV}$.

Note that the value of the exchange interaction between the oxygen
subsystem and the subsystem of the localized spins calculated
using the expression (\ref{Hamiltonian_notations}) is turned out
to be large: $J=3.4\,\textrm{eV}\gg \tau \approx0.1\,\textrm{eV}$.
Thus, the dynamics of the oxygen holes should be described taking
into account their correlation with the subsystem of spins of
copper ions. This problem can be solved using the basis set of
operators~\cite{Dzebisashvili13,Val'kov14}
\begin{equation}\label{spin-fermion_basis_normal}
a_{k\alpha},\quad b_{k\alpha},\quad
L_{k\alpha}=\frac1N\sum_{fq\beta} e^{if(q-k)}
(\textbf{S}_f\boldsymbol{\sigma}_{\alpha\beta})u_{q\beta},
\end{equation}
where the third operator couples both the spin and the fermion
dynamics.

\vspace{8pt}

{\bf 3. Equations for Green's functions}

To consider the conditions for the Cooper instability, let us add
the operators ($\bar{\alpha}=-\alpha$)
\begin{eqnarray}\label{spin-fermion_basis_supercond}
a_{-k\bar{\alpha}}^{\dag},\quad b_{-k\bar{\alpha}}^{\dag},\quad
L_{-k\bar{\alpha}}^{\dag}
\end{eqnarray}
to the basis set (\ref{spin-fermion_basis_normal}). The system of
equations for the normal $G_{ij}$ and the anomalous $F_{ij}$
Green's functions obtained in the framework of the method
\cite{Zwanzig61,Mori65} can be written as follows ($j=1,2,3$)
\begin{eqnarray}\label{equations}
&&(\omega-\xi_{x})G_{1j} = \delta_{1j} + t_{k}G_{2j}+J_{x}G_{3j}
+\Delta_{1k}F_{2j},\nonumber\\
&&(\omega-\xi_{y})G_{2j} = \delta_{2j}+t_{k}G_{1j}+J_{y}G_{3j}
+\Delta_{2k}F_{1j},\nonumber\\
&&(\omega-\xi_{3})G_{3j} = \delta_{3j}K_{k}+
(J_{x}G_{1j}+J_{y}G_{2j})K_{k}
+\Delta_{3k}F_{3j},\nonumber\\
&&(\omega+\xi_{x})F_{1j} = \Delta_{2k}^*G_{2j}
-t_{k}F_{2j}-J_{x}F_{3j},\nonumber\\
&&(\omega+\xi_{y})F_{2j} = \Delta_{1k}^*G_{1j}-
t_{k}F_{1j}-J_{y}F_{3j},\nonumber\\
&&(\omega+\xi_{3})F_{3j} =
\Delta^*_{3k}G_{3j}-(J_{x}F_{1j}+J_{y}F_{2j})K_{k}.
\end{eqnarray}
Here the defenitions
\begin{eqnarray*}
G_{11}=\langle\langle
a_{k\uparrow}|a_{k\uparrow}^{\dag}\rangle\rangle,~
G_{21}=\langle\langle
b_{k\uparrow}|a_{k\uparrow}^{\dag}\rangle\rangle,~
G_{31}=\langle\langle
L_{k\uparrow}|a_{k\uparrow}^{\dag}\rangle\rangle.
\end{eqnarray*}
are used. The functions $G_{i2}$ and $G_{i3}$ are defined
similarly, with the difference that instead $a^{\dag}_{k\uparrow}$
the operators $b^{\dag}_{k\uparrow}$ and $L^{\dag}_{k\uparrow}$
stand, respectively. The anomalous Green's functions are
\begin{eqnarray*}
F_{11}=\langle\langle
a_{-k\downarrow}^{\dag}|a_{k\uparrow}^{\dag}\rangle\rangle,\,
F_{21}=\langle\langle
b_{-k\downarrow}^{\dag}|a_{k\uparrow}^{\dag}\rangle\rangle,\,
F_{31}=\langle\langle
L_{-k\downarrow}^{\dag}|a_{k\uparrow}^{\dag}\rangle\rangle.
\end{eqnarray*}
At that, for $F_{i2}$ and $F_{i3}$ the same definitions regarding
the second index are used. The functions in (\ref{equations}) are
\begin{eqnarray}\label{notations}
&&\xi_{x(y)}=\xi_0({k_{x(y)}})+4n_pV,\nonumber\\
&&J_{x(y)}=J\cos\frac{k_{x(y)}}{2},\quad K_{k}=3/4+C_1\gamma_{1k},\nonumber\\
&&\xi_{3}=\varepsilon_p-\mu-2t+5\tau/2-J+n_pV+\nonumber\\
&&\quad+\left[(\tau-2t)(C_1\gamma_{1k}+C_2\gamma_{2k})
+\tau(C_1\gamma_{1k}+C_3\gamma_{3k})/2\right.+\nonumber\\
&&\quad\left.+JC_1(1-4\gamma_{1k})/4+IC_1(\gamma_{1k}-4)\right]K_{k}^{-1}.
\end{eqnarray}
Here, the number of holes per one oxygen ion is $n_p$,
$\gamma_{jk}$ are the square lattice invariants:
$\gamma_{1k}=(\cos k_x+\cos k_y)/2,\quad \gamma_{2k}=\cos
k_x\,\cos k_y,\quad\gamma_{3k}=(\cos 2k_x+\cos 2k_y)/2$. To obtain
(\ref{equations}), we assume that the subsystem of the localized
spins is in the quantum spin liquid state. In this case, the spin
correlation functions $C_j=\langle
\textbf{S}_0\textbf{S}_{r_j}\rangle$ satisfy the relations
\begin{equation}\label{spin_correlators}
C_j=3\langle S^x_0S^x_{r_j}\rangle=3\langle
S^y_0S^y_{r_j}\rangle=3\langle S^z_0S^z_{r_j}\rangle,
\end{equation}
where $r_j$ is a coordinate of the copper ion in the coordination
sphere $j$. At that, $\langle S^x_f\rangle=\langle
S^y_f\rangle=\langle S^z_f\rangle=0$.

Using system (\ref{equations}), one can obtain that the fermionic
spectrum in the normal phase can be obtained from the dispersion
equation
\begin{eqnarray}\label{det}
&&\mathrm{det}_{k}(\omega)=(\omega-\xi_{x})(\omega-\xi_{y})
(\omega-\xi_{3})-2J_{x}J_{y}t_{k}K_{k}-\nonumber\\
&&-(\omega-\xi_{y})J_{x}^2K_{k} -(\omega-\xi_{x})J_{y}^2K_{k}
-(\omega-\xi_{3})t_{k}^2=0.
\end{eqnarray}
The spectrum consists of the three branches $\epsilon_{1k}$,
$\epsilon_{2k}$ and $\epsilon_{3k}$~\cite{Val'kov15}. The
appearing of the branch $\epsilon_{1k}$ with the minimum near
($\pi/2$, $\pi/2$) is caused by the strong spin-fermionic
correlation which initiates both the exchange interaction between
the hole and the nearest copper ions and the spin-correlated
hoppings. At the low number $n_p$, the dynamics of holes is mainly
defined by the lower band $\epsilon_{1k}$, which is significantly
separated from the upper bands $\epsilon_{2k}$ and
$\epsilon_{3k}$~\cite{Val'kov15}.


The superconducting order parameters $\Delta_{j,k}$ are connected
with the anomalous averages by means the following expressions
($C_{1x(1y)}=C_1\cos^2 (q_{x(y)}/2)$):
\begin{eqnarray}\label{Deltas}
&&\Delta_{1k}=-\frac{4V}{N}\sum_{q}\phi_{k-q}\langle
a_{q\uparrow}b_{-q\downarrow}\rangle,~~~
\phi_{k}=\displaystyle\cos\frac{k_x}{2}\cos\frac{k_y}{2},
\nonumber \\
&&\Delta_{2k}=-\frac{4V}{N}\sum_{q}\phi_{k-q}\langle
b_{q\uparrow}a_{-q\downarrow}\rangle,\\
\label{Delta3I} &&\Delta_{3k}=\frac
1N\sum_{q}\biggl\{I_{k-q}\Bigl[\langle
L_{q\uparrow}L_{-q\downarrow}\rangle -\nonumber\\
&&\quad-C_{1x}\langle
a_{q\uparrow}a_{-q\downarrow}\rangle-C_{1y}\langle
b_{q\uparrow}b_{-q\downarrow}\rangle\Bigr]K_k^{-1}+\nonumber\\
&&\quad+(\widetilde{V}_k-C_1I_{k-q}K_k^{-1})\phi_{q} \bigl(\langle
a_{q\uparrow}b_{-q\downarrow}\rangle+\langle
b_{q\uparrow}a_{-q\downarrow}\rangle\bigr)\biggr\}.\nonumber
\end{eqnarray}
Here $\widetilde{V}_k=V\left(1+(C_1\gamma_{1k}
+C_2\gamma_{2k})K_k^{-1}\right)$.

\vspace{6pt}

\textbf{4. The system of equations for the superconducting order
parameters}

To analyze the conditions for the Cooper instability, let us
express in the linear approximation the anomalous Green's
functions in terms of the parameters $\Delta^*_{jk}$
\begin{eqnarray}\label{Fij}
&&F_{nm}(k,\omega)=\sum_{j=1}^3S^{(j)}_{nm}(k,\omega)\Delta_{jk}^*/\textrm{Det}_k(\omega),
\end{eqnarray}
where the following notations are used:
\begin{eqnarray}
&&\textrm{Det}_k(\omega)=-\textrm{det}_k(\omega)\textrm{det}_k(-\omega),\nonumber\\
&&S^{(1)}_{12}(k,\omega)~=~S^{(2)}_{21}(k,\omega)~=~ Q_{3}(k,-\omega)Q_{3}(k,\omega),\nonumber\\
&&S^{(2)}_{12}(k,\omega)=Q_{3y}(k,-\omega)Q_{3x}(k,\omega),\nonumber\\
&&S^{(3)}_{12}(k,\omega)=K_kQ_{y}(k,-\omega)Q_{x}(k,\omega),\nonumber\\
&&S^{(1)}_{21}(k,\omega)~=~S^{(2)}_{12}(k,-\omega),~~
 S^{(3)}_{21}(k,\omega)~=~S^{(3)}_{12}(k,-\omega),\nonumber\\
&&S^{(1)}_{11}(k,\omega)~=~Q_{3}(k,-\omega)Q_{3y}(k,\omega),\nonumber\\
&&S^{(2)}_{33}(k,\omega)=K_kS^{(3)}_{12}(k,\omega),~
 S^{(2)}_{11}(k,\omega)~=~S^{(1)}_{11}(k,-\omega),\nonumber\\
&&S^{(3)}_{11}(k,\omega)=K_kQ_{y}(k,-\omega)Q_{y}(k,\omega),\nonumber\\
&&S^{(1)}_{22}(k,\omega)=Q_{3x}(k,-\omega)Q_{3}(k,\omega),\nonumber\\
&&S^{(1)}_{33}(k,\omega)=K_kS^{(3)}_{21}(k,\omega),~S^{(2)}_{22}(k,\omega)=S^{(1)}_{22}(k,-\omega),\nonumber\\
&&S^{(3)}_{22}(k,\omega)=K_kQ_{x}(k,-\omega)Q_{x}(k,\omega),\nonumber\\
&&S^{(3)}_{33}(k,\omega)=K_kQ_{xy}(k,-\omega)Q_{xy}(k,\omega).\nonumber
\end{eqnarray}
These expressions include the functions
\begin{eqnarray}
&&Q_{x(y)}(k,\omega)=(\omega-\xi_{x(y)})J_{y(x)}+t_kJ_{x(y)},\nonumber\\
&&Q_3(k,\omega)=(\omega-\xi_3)t_k+J_xJ_yK_k,\nonumber\\
&&Q_{3x(3y)}(k,\omega)=(\omega-\xi_3)(\omega-\xi_{x(y)})-J_{x(y)}^2K_k,\nonumber\\
&&Q_{xy}(k,\omega)=(\omega-\xi_x)(\omega-\xi_y)-t_k^2.
\end{eqnarray}
Using the spectral theorem~\cite{Zubarev60}, we obtain the
expressions for the anomalous averages and the closed system of
homogenous integral equations for the superconducting order
parameters
\begin{eqnarray}\label{Deltas_spectral_theorem}
&&\Delta_{1k}^*=\frac {4V}{N}\sum_{jq}\phi_{k-q}M^{(j)}_{21}(q)
\Delta_{jq}^*,\nonumber\\
&&\Delta_{2k}^*=\frac {4V}{N}\sum_{jq}\phi_{k-q}M^{(j)}_{12}(q)
\Delta_{jq}^*,\nonumber\\
&&\Delta_{3k}^*=\frac{1}{N}\sum_{jq}\left\{\frac{I_{k-q}}{K_k}\left[
C_{1x}M^{(j)}_{11}(q)+C_{1y}M^{(j)}_{22}(q)-M^{(j)}_{33}(q)\right]\right.\nonumber\\
&&\qquad\left.+(I_{k-q}C_1K_k^{-1}-\widetilde{V}_k)\phi_q\left[M^{(j)}_{12}(q)+M^{(j)}_{21}(q)\right]\right\}\Delta_{jq}^*,
\end{eqnarray}
where
$$ M^{(j)}_{nm}(q)=\frac{S^{(j)}_{nm}(q,E_{1q})
+S^{(j)}_{nm}(q,-E_{1q})}{4E_{1q}(E_{1q}^2-E_{2q}^2)
(E_{1q}^2-E_{3q}^2)}\tanh\left(\frac{E_{1q}}{2T}\right).
$$

Below, we use the system (\ref{Deltas_spectral_theorem}) to find
the critical temperature of the transition to the superconducting
phase with the assigned type of symmetry of the order parameter.
\vspace{10pt}

\textbf{5. The critical temperature of the superconducting
$d_{x^2-y^2}$-wave pairing}

For the superconducting $d_{x^2-y^2}$-wave pairing, when
\begin{eqnarray}\label{Dcos}
\Delta_{3k}=\Delta_0\cdot(\cos k_x-\cos k_y),
\end{eqnarray}
it follows from the system (\ref{Deltas_spectral_theorem}) that
$\Delta_{1k}=0$ and $\Delta_{2k}=0$. It is easily seen if one
takes into account that the kernels of the integral equations for
$\Delta^*_{1k}$ and $\Delta^*_{2k}$ contain the function
$\phi_{k-q}$. As a result, the integration over $q$ vanishes these
superconducting order parameters.

It follows from the integral equation for the third
superconducting order parameter $\Delta^*_{3k}$ that the
contribution of the intersite Coulomb potential to the kernel of
the integral equation is equal to zero. This is due to the
symmetry properties of the integrands and manifests itself after
summation over the internal variable. As a result, we find that
the Coulomb repulsion between the holes located at the nearest
oxygen ions does not suppress the superconducting
$d_{x^2-y^2}$-wave pairing. Thus, we arrive at the equation for
the concentration dependence of the critical temperature
\begin{eqnarray}\label{Tc_eq}
1=\frac {I}{N}\sum_{q}\frac{(\cos q_x-\cos q_y)^2}
{2E_{1q}}\Psi_q\tanh\biggl(\frac{E_{1q}}{2T_c}\biggr),
\end{eqnarray}
where
\begin{eqnarray}
&&\Psi_q=\biggl\{S_{33}^{(3)}(E_{1q})-C_{1x}S_{11}^{(3)}
(E_{1q})-C_{1y}S_{22}^{(3)}(E_{1q})-\nonumber\\
&&\quad-C_1\phi_q\left(S_{12}^{(3)}(E_{1q})+S_{21}^{(3)}
(E_{1q})\right)\Bigr]\biggr\}\times\\
&&\quad\times
\left[K_q(E_{1q}^2-E_{2q}^2)(E_{1q}^2-E_{3q}^2)\right]^{-1}.\nonumber
\end{eqnarray}

\begin{figure}[t]
\begin{center}
\includegraphics[width=0.4\textwidth]{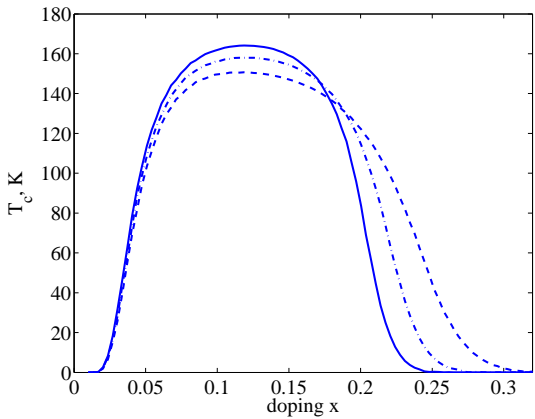}
\caption{Fig. 1. Concentration dependence of the critical
temperature of the transition to the superconducting
$d_{x^2-y^2}$-wave phase calculated for $V=0$ (dotted curve),
$V=1\,\textrm{eV}$ (dashed-dotted curve) and $V=2\,\textrm{eV}$
(solid curve).}\label{Tcn}
\end{center}
\end{figure}

Figure~\ref{Tcn} shows the dependencies of the superconducting
critical temperature on doping obtained by solving Eq.
(\ref{Tc_eq}). A comparison of the curves shows that an account
for the intersite Coulomb repulsion leads to insignificant and
nonuniform in doping modification of the dependence $T_c(n)$. Note
that these insignificant modifications are caused by the
renormalization of the one-site energy of holes owing to the
intersite Coulomb repulsion, but not the renormalization of the
coupling constant.

\vspace{10pt}

{\bf 4. Conclusion}

The main result of the paper is connected with the answer to the
question of why the superconducting $d$-wave pairing survives with
regard to the intersite Coulomb repulsion between oxygen holes, as
well as why the $d_{x^2-y^2}$-wave pairing instead of the $s$-wave
pairing occurs in cuprate superconductors despite the strong
coupling constant which corresponds to the kinematic mechanism.
For the analysis of the conditions for the Cooper instability in
cuprate superconductors in the framework of the exchange,
kinematic and spin-fluctuation mechanisms, the effective models
(the Hubbard model, $t-J$, $t-J^*$ models) on the lattice with a
primitive cell were mainly used. An account for the intersite
Coulomb interaction in these models led to the suppression of the
$d_{x^2-y^2}$-wave pairing, whereas the superconducting $s$-wave
pairing initiated by the kinematic mechanism survived. As a
result, the contradiction between theoretical and experimental
results arised: the experiment demonstrated the superconducting
$d_{x^2-y^2}$-wave pairing, whereas theoretically this pairing has
been suppressed.

We established that the key to the resolving of the
above-mentioned contradiction is connected with an account for the
real structure of CuO$_2$ plane. It appears to be that the Fourier
transform of the Coulomb potential slips out of the system of the
integral equations for the superconducting order parameters, as
soon as the solution corresponding to the $d_{x^2-y^2}$-wave
pairing is considered. Therefore, the Coulomb repulsion between
holes located at the nearest oxygen ions does not suppress the
Cooper pairing in the $d$-wave channel. And conversely, the
equation for the $s$-wave pairing contains the Coulomb potential
which leads to the suppression of superconductivity. Note that the
different contributions of the Coulomb repulsion to the conditions
of realization of the superconducting phases with the different
types of the symmetry of the order parameter also manifest itself
in the Kohn-Luttinger theory of superconductivity~\cite{Kagan15}.
In our case, the space separatedness of the two-orbital subsystem
of the oxygen holes and the subsystem of the localized spins of
copper ions plays a leading role. It is now apparent that the
theories based on the models which use the lattices with a
primitive cell, instead of the real structure, are inappropriate
for realistic theoretical consideration of the properties of
cuprate superconductors.

In conclusion, let us dwell on the uncovered property of symmetry
which leads to the absence of the contribution of the Coulomb
repulsion between the nearest oxygen holes to the $d$-wave
pairing. In traditional superconductors, the contribution of the
Coulomb potential is renormalized due to the electron-phonon
interaction, whereas in high-temperature superconductors the
neutralization of the Coulomb repulsion for the $d_{x^2-y^2}$-wave
pairing is due to a non-primitive unit cell and a specific
character of the Fourier transform of the Coulomb potential.
Hence, an important principle is emerges that allows to realize
the goal-oriented search of new high-T$_c$ superconducting
systems. Such systems should have the lattice with a non-primitive
unit cell and the lattice should possess the structure for which
the contribution of the Fourier transform of the intersite Coulomb
interaction to the integral equation for the superconducting gap
vanishes. This is the situation that occurs in cuprate
superconductors.

The work was supported by the Russian Foundation for Basic
Research (project nos. 16-02-00073 and 16-02-00304). The work of
D.\,D.\,M. and M.\,M.\,K. was supported by the Dynasty Foundation.

\vfill\eject

\end{document}